\documentstyle[12pt]{article}
\font\tenbf=cmbx10
\font\tenrm=cmr10
\font\tenit=cmti10
\font\elevenbf=cmbx10 scaled\magstep 1
\font\elevenrm=cmr10 scaled\magstep 1
\font\elevenit=cmti10 scaled\magstep 1

\begin{document}
\footnotetext{\it To appear in Proceedings of DPF '92: Meeting of the American
Physical Society (Fermilab 1992) (World Scientific, in press)}

\begin{center}{{\tenbf HIGH ENERGY HADRONIC HELICITY VIOLATION:\\
               \vglue 3pt
               WHAT HELICITY FLIP FORM FACTORS MEASURE \\}
\vglue 5pt

{\tenrm PANKAJ JAIN and JOHN P. RALSTON\\}
\baselineskip=13pt
{\tenit Department of Physics and Astronomy, University of Kansas \\}
\baselineskip=12pt
{\tenit Lawrence KS, 66045, USA\\}
\vglue 0.3cm
{\tenrm and\\}
\vglue 0.3cm
{\tenrm BERNARD PIRE\\}
{\tenit Centre de Physique Theorique, Ecole Polytechnique, F91128, Palaiseau,
France\\}
\vglue 0.8cm
{\tenrm ABSTRACT}}
\end{center}
\vglue 0.3cm
{\rightskip=3pc
 \leftskip=3pc
 \tenrm\baselineskip=12pt
 \noindent Orbital angular momentum is crucial for the description of
several process at large momentum transfer in the impulse approximation
in perturbative QCD.  We review the mechanism of independent scattering
which makes violation of the helicity conservation rule possible at
leading order - and without any need to flip a quark helicity - in exclusive
hadron-hadron reactions.  We argue that helicity flip form factors,  which
violate the factorization assumptions of conventional approaches and are
classified as ``higher-twist," are nevertheless calculable to leading order
in perturbative QCD when information on non-zero orbital angular
momentum is incorporated.

\vglue 0.6cm}
\vglue 0.1cm
\baselineskip=14pt
\elevenrm
For a long time the hadronic helicity conservation rule,$^1$
$$\lambda_A+\lambda_B=\lambda_C+\lambda_D\eqno (1)$$ for reactions of hadrons
(A--D) with helicities $\lambda_i$,
has been considered to be a firm
prediction of perturbative QCD.  The rule is important because it is very
general, as general as the factorization of the quark-counting procedure.
The fact that this rule is badly violated in almost every case in which it
has been experimentally checked suggests two alternatives:  that the
momentum transfer ($Q^2$) is not high enough, or that the factorization is
not satisfied. Theorists have been divided on these options.  The factorization
is
theoretically attractive, and there are several impressive experimental
successes of
the quark-counting rules, indicating that a few pointlike constituents are
being observed in hard scattering.  On the other hand, several pieces of
data show large violations of hadronic helicity conservation which do not
seem to get smaller with increasing $Q^2$.

	Is there a way out?  Recently it was shown that the {\em independent
scattering} graphs generally break the assumptions needed to obtain
hadronic helicity conservation$^2$.  The new insight is that independent
scattering$^3$ occurs over an interaction region which is not ``small and
round" as assumed in the quark counting rules, but rather a flattened
spatial area correlated with the scattering plane.  If the hard scattering
exchanges momentum lying in the scattering plane (y) direction, then there is a
short distance of spatial separation $<b_y^2> = 1/Q^2$,  but the out-of-plane
separation  $<b_x^2>$ is not so small and is set by the integrations over the
hadron wave functions.  In contrast to the factorization of the quark-counting
process, which to leading order imposes a sort of filter selecting
quarks having zero orbital angular momentum, {\em all}  possible orbital
angular
momenta can participate in independent scattering.  Once a unit of orbital
angular momentum enters, hadrons can flip their helicity through the wave
functions, which are non-perturbative and part of the ``long-time"
evolution the impulse approximation uses as input.  There is no need to flip a
quark helicity; quark mass terms can be dropped.
Hadron helicity
{\em non}-conservation is therefore a simple consequence of looking at the
ordinary QCD
description of independent scattering process diagrams in a real-space
coordinate system.  It has gone unrecognized for a long time, because
little was done in coordinate space until the work of Botts and
Sterman$^4$, who also described a systematic procedure for calculating the
hard scattering.

	 The impulse approximation can still be used in the limit $Gev^2 < Q^2
<TeV^2$, where Sudakov effects provide a fascinating ``pre-confining"
perturbative cut-off of dangerously large transverse space separations
between quarks.  The spatial separation amplitudes follow the rule of
``survival of the smallest," but this effect is not rapid enough to suppress
the independent scattering diagrams altogether; instead, it tends to
suppress the potentially non-perturbative contributions.  Botts$^4$ showed
with numerical studies that the asymptotic Sudakov suppression, long
argued to be a good reason to ignore the independent scatterers, set in
only for $Q^2 > TeV^2$.  Thus the independent scattering contributions are
indicated by every approach: theory,  energy dependence, and spin dependence.

	To calculate further, information must be obtained on the non-zero
orbital angular momentum wave functions.  We suggest$^5$ studying
helicity-flip form factors for this, again assuming that quark mass
effects can be dropped - an excellent approximation if the relevant quark
mass is really a few MeV. While a large-$Q^2$ photon probe does select short
distance parts of the wave functions,  the quark counting factorization
can not be applied naively, because it says helicity violation is forbidden.
We analyze the process in terms of the quark-hadron scattering amplitude,
with a hadron helicity flip.  This can be non-zero in our impulse
approximation, and indeed it is non-zero (as we claimed above) when
there is independent scattering.   The quark hadron scattering measures
non-zero orbital angular momentum;  pinching the ends of the scattering
shut with the insertion of a hard probe,  we have the helicity flip form
factor.  In our power counting of this process a wave function carrying
orbital angular momentum $m$  goes like $b^m$ as $b \to 0$; each power of $b$
in the wave function probed with a good form factor scales like $1/Q$. The
overall appearance of the form factor $F(Q^2)$ after removing kinematic
effects is basically
$$F(Q^2) = \int d^2k_T\psi^*(k_T+Q)\psi(k_T)=\sum\limits_{mm'}\int
d^2b_T\tilde\psi_{m'}^*(b_T)e^{iQ_T\cdot b_T}\tilde\psi_m(b_T)\eqno (2)$$
where in the second equation we made an orbital angular momentum
expansion.  The proton's $F_2(Q^2)$ form factor, for example, is then found to
go like $1/Q^6$ due to the presence of m = 1 and to be entirely
{\em calculable} in
leading order perturbative QCD.  Space does not permit details, which will be
presented elsewhere:  it requires a long series of definitions to say
exactly which wave functions are being measured.   We can assert that
recent SLAC data measuring $F_2$ and conforming to the power counting
allows us to conclude that $m = 1$ wave functions in the proton are not
anomalously small compared to the $m = 0$ parts.

	The spin structure of hadrons is growing more and more interesting.
The idea that $F_2$  measures non-zero orbital angular momentum can be
tested in color transparency:  electroproduction knockout of protons (or
the self-analyzing deltas)  from a nuclear target,  with measured final
state polarization.  Non-zero m should be preferentially filtered away at
large A.   It turns out that the new non-short distance wave functions,
which we have suggested measuring with short distance probes, have
something deeply in common with the chirally odd inclusive correlation
functions that have generated much interest recently$^6$.  The chirality of
a wave function-  defined by its commutation relations with gamma-5 -
measures the helicity orientation, which is necessarily coupled to the
orbital angular momentum.  We expect more theoretical work in this area,
and exciting interaction with experiment.

\vglue 0.5cm
{\elevenbf \noindent Acknowledgements\hfil}
\vglue 0.4cm
This work was supported in part under DOE grant
number DE-FG02 85-ER 40214.A0009.

\vglue 0.5cm
{\elevenbf\noindent References\hfil}
\vglue 0.4cm

\begin{enumerate}
\item S. J Brodsky and G. P. LePage, {\elevenit Phys Rev} {\elevenbf D22},
2157(1980);  G. P. Lepage
and S. J. Brodsky, {\elevenit Phys Rev.} {\elevenbf D22}, 2157,(1980).

\item J. P. Ralston and B. Pire, ``High Energy Helicity Violation in Hard
Exclusive Scattering of Hadrons"  Kansas and Ecole Polytechnique preprint
A175.0592, submitted to {\elevenit Phys. Lett B}. See also in {\elevenit
Polarized Collider Workshop} (Penn State, 1990) ed. by J. Collins et al. (AIP
Conference Proceedings No. 223) p. 228.

\item P. V. Landshoff, {\elevenit Phys Rev} {\elevenit D10}, 1024 (1974).

\item J. Botts and G. Sterman, {\elevenit Nuc Phys} {\elevenbf B325}, 62
(1989);  J. Botts,
ibid  {\elevenbf 353}, 20,(1989).

\item in preparation

\item J. L. Cortes, B. Pire and J. P. Ralston, in {\elevenit Polarized Collider
Workshop}
(Penn State, 1990)  edited by J. C. Collins et al ( APS \#223) p 176, and in
{\elevenit Zeit Phys} {\elevenbf C55} 409(1992); R.L Jaffe and X. Ji,
{\elevenit Phys Rev Lett}
{\elevenbf 67}, 552, (1991).
\end{enumerate}
\end{document}